\documentstyle[prl,aps,epsfig,floats]{revtex}

\title {Liquid-like spatial distribution of magnetic polarons revealed by neutron scattering in La$_{1-x}$Ca$_{x}$MnO$_3$}
\author{M. Hennion$^1$, F. Moussa$^1$, G. Biotteau$^1$, J. Rodr\'{\i}guez-Carvajal$^1$,
L. Pinsard$^2$ and A. Revcolevschi$^2$ }

\address{1 - Laboratoire L\'eon Brillouin, CEA-CNRS, CE Saclay, 
91191 Gif sur Yvette Cedex France}
\address{2 - Laboratoire de Chimie des Solides 
Universit\'e Paris-Sud 91405 Orsay Cedex France}
\date{\today, submitted to Phys. Rev. Lett.}
\begin{document}
\twocolumn[\hsize\textwidth\columnwidth\hsize\csname @twocolumnfalse\endcsname
\maketitle
\begin{abstract}
Elastic neutron scattering experiments performed in  semi-conducting La$_{1-x}$Ca$_x$MnO$_3$ single crystals (x=0.05, 0.08), reveal new features in the problem of electronic phase separation and metal insulator transition. Below T$_N$, the observation of a broad magnetic modulation in the q-dependent scattering intensity, centered at nearly identical q$_m$ whatever the q direction, can be explained by a liquid-like spatial distribution of 
 similar magnetic droplets. A semi-quantitative description of their magnetic state, diameter, and average distance , can be done  using a two-phase 
model. Such a picture can explain the anomalous characteristics of the spin wave branches and may result from 
unmixing forces between charge carriers predicted from the s-d model.
 \end{abstract}
]

Doped Mn perovskites are intensively studied for their remarkable giant magnetoresistance underlying the electronic behaviour that has potential technological applications. The basic physical ideas have been given by Zener\cite{Zener}, with the model of the double exchange, which connects the electron hopping with the ferromagnetic (F) alignments of Mn spins. The electron-phonon coupling through the Jahn-Teller effect was studied by Millis et al.\cite{Millis}. On the other hand, the r\^ole of the electronic localization and the formation of magnetic polarons in this problem has been emphasized by Varma et al.\cite{Varma}. Several pictures of magnetic polarons were proposed, associated to one carrier (small polarons)\cite{de Gennes,Kasuya} or, in the case of larger carrier densities, to several carriers (large polarons)\cite{Nagaev}. A cooperative state of magnetic droplets by self-trapping carriers was predicted by Nagaev using the s-d or s-f model\cite{Nagaev}, since this physical situation was first suggested for some rare earth compounds\cite{Oliveira}. Models of electronic phase diagram have recently been reexamined\cite{Dagotto,Khomskii}. Magnetic inhomogeneities have been suggested to be at the origin of the behaviour of the susceptibility and of the dynamical spin fluctuations observed above T$_c$ in doped Mn perovskites\cite{Lynn,De Teresa}. Very recent NMR experiments have been interpreted in terms of electronic phase separation\cite{Allodi}. However, a characterization of magnetic inhomogeneities has never been made until now. We have 
started a spin dynamics study in La$_{1-x}$Ca$_x$MnO$_3$ for low 
doping\cite{Moussa,Hennion} where the system is insulating and undergoes a weakly canted antiferromagnetic (AF) transition at T$_N$\cite{Wollan}. An additional spin wave branch occurs by doping. The corresponding dynamical susceptibility reveals a small ferromagnetic correlation length, whereas the dispersion
curve is isotropic, indicating a new ferromagnetic coupling between some Mn spins. This isotropy contrasts with the strongly anisotropic character of the other spin wave branch, characteristic of super-exchange. This new spin dynamics was attributed to "magnetic polarons", whose origin could not be elucidated. 
  
 We have performed {\it elastic} neutron scattering measurements as a function of temperature in La$_{1-x}$Ca$_x$MnO$_3$ (x=0.05, 0.08), close to the direct beam ($\tau=0$) and to $\tau$=(110) Bragg peak. In both 
experimental cases, a broad modulation is observed, centered 
at nearly the same q$_m$ whatever the q direction, with intensity growing below T$_N$. This scattering does not exist for x=0. Such a pattern is typical of an assembly of magnetic clusters or droplets,  with a mean magnetization different from that of the matrix and a well-defined shortest distance between them. A semi-quantitative description of the droplets and of their spatial distribution is proposed using a liquid-like model. The droplets are isotropic, their radius is $\approx 9\AA$, their density is low compared to that of the hole concentration and their minimal distance of approach indicates a repulsive interaction.The existence of these magnetic clusters is closely linked with the unusual spin dynamics\cite{Hennion,Moussa2,Hennion2}.      
The overall static and dynamic observations suggest a picture of magnetic droplets or large polarons with a magnetic coupling distinct from exchange, coupled through the surrounding medium characterized by 
the super-exchange coupling.  They are characteristic of an electronic phase segregation, which results from the balance between unmixing forces, predicted 
by the s-d model, and intermixing Coulomb forces.
 \begin{figure}
\centerline{\epsfig{file=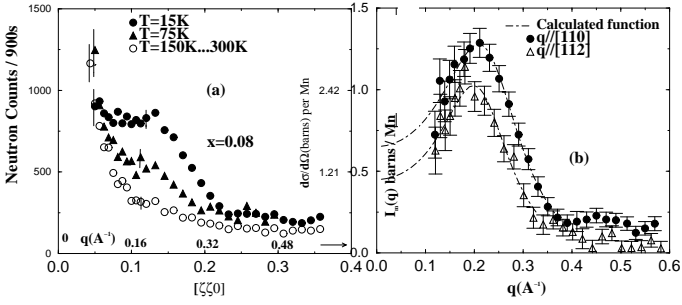,width=9cm}}
\caption{{\bf a)} Scattering Intensities versus q, calibrated in barns, observed above T$_N$  (I(q)=I(q)$_{HT}$), at 75K and 15K for [110]$_1$ q direction. {\bf b)} Magnetic intensities I$_m$(q)=I(q)-I$_{HT}$(q) versus q at 15K for [110]$_1$, and  [112]$_1$ q directions. The dashed lines  are calculated functions according to the model described in the text.}
\label{fig:35_6.8} \end{figure}
\begin{figure}
\centerline{\epsfig{file=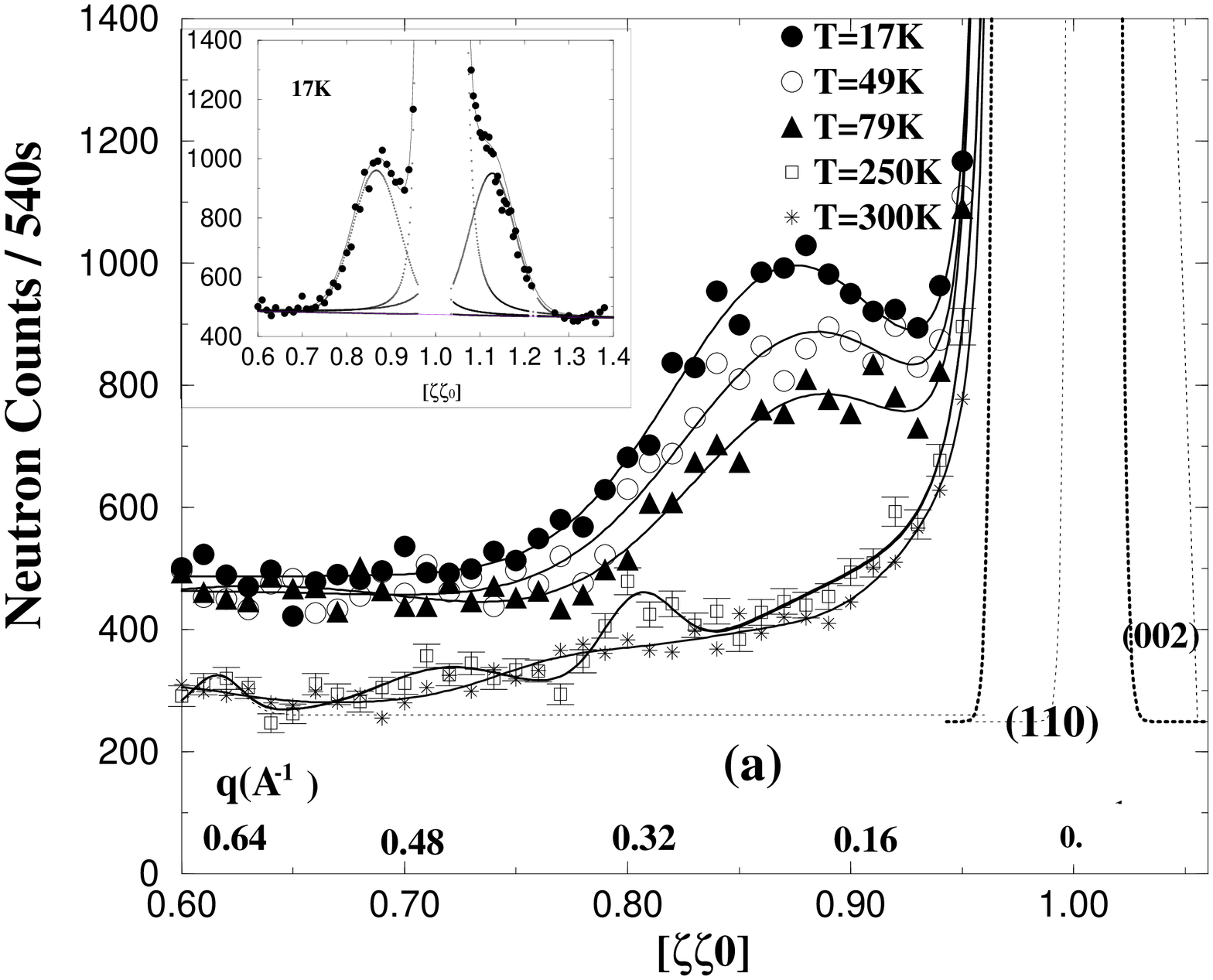,width=8cm}}
\centerline{\epsfig{file=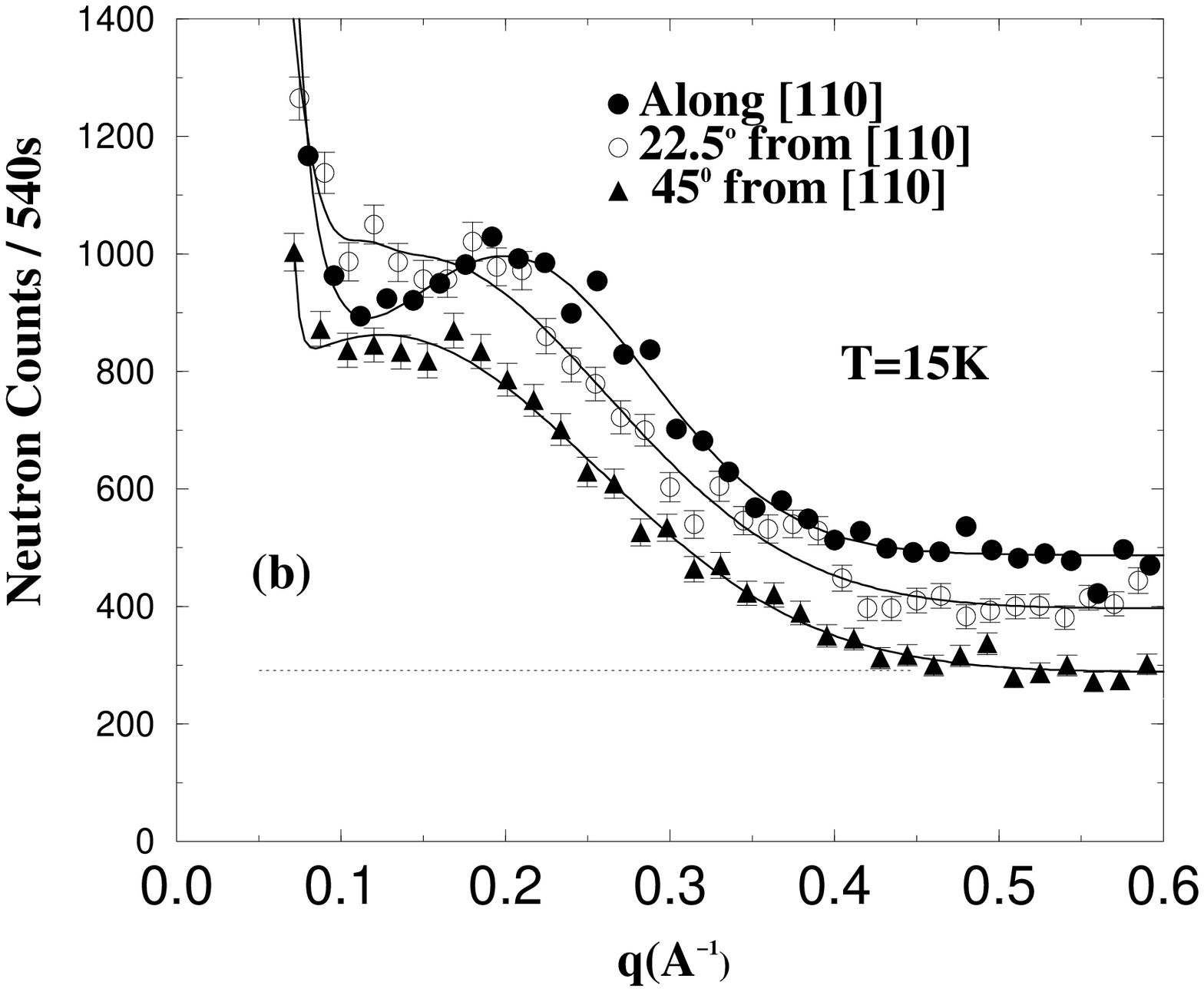,width=8.2cm}}
\caption{Diffuse scattering intensities versus Q (or q=Q-$\tau$) measured around $\tau$=(110)$_1$. {\bf a)} along [110]$_1$, at five temperatures in the 300K-17K range (T$_N$=122K). When not shown, the error bar is within the size of the symbol. In the inset, 
 the continuous lines are components of a fit with two (+q$_m$, -q$_m$) modulations, symmetric with respect to (110)$_1$. 
 {\bf b)} At 15K along three q directions. The horizontal line locates the sample background observed at 300K (cf fig 2-a).}.
\label{fig:35_6.8}
 \end{figure}

 Experiments have been 
carried out at the reactor
 Orph\'ee (Laboratoire L\'eon Brillouin) using a triple axis spectrometer set on a cold neutron 
source. The samples with x=0.05 and 0.08 show similar results, so that only observations for the x=0.08 sample, with a volume of 0.3 cm$^3$ are reported.
The crystalline structure is orthorhombic (O') with {\it Pbnm} symmetry. The crystals are twinned basically having three domains of volume V$_i$ ($i$=1,2,3) with their {\bf c} axes along the 
directions {\bf a$_0$, b$_0$, c$_0$} of the cubic perovskite substructure (called F domains in\cite{Moussa}). The mirror or M domains, leading to two peaks in the rocking curve, may have some influence close to $\tau \ne 0$. The {\bf q} directions [110], [001],[112] and [100] refers to  {\bf a$_0$, c$_0$}, to the diagonals of the faces ({\bf a$_0$+c$_0$}) and ({\bf a$_0$+ b$_0$}) respectively. The subscript $i$ will be used below to distinguish the {\bf q} direction within the different F domains. The magnetic structure has been determined from Bragg peak intensities. It consists of ferromagnetic ({\bf a-b}) planes, with AF stacking along  {\bf c}. However, the increase of the (110) and (112) nuclear Bragg peaks below T$_N$ (122K), indicates a long range ferromagnetic ordering for a small spin component along  {\bf c}. It corresponds to a canted state. At 14K, the {\it average} canting angle, $\theta_{av}$, is $\approx$$10$ degrees from {\bf b} within the ({\bf b-c}) plane.

{\it Elastic} experiments have been performed around $\tau=0$ with k$_i$=1.25$\AA^{-1}$, covering 6 temperatures within the 15K$<T<$300K and 0.05$<q<$0.6$\AA{-1}$ ranges, along several {\bf q} directions. Intensities have been put on an absolute scale, using a vanadium sample and a standard procedure for correction. The temperature dependence of the spectra are reported in Fig 1-a for q//[110]$_1$. In the $150K<T<300K$ range, I(q)= I(q)$_{HT}$ is found temperature independent. In the smallest q range, I(q)$_{HT}$ is attributed to dislocations or large structural defects present in the sample. At larger q, the residual intensity is nearly twice as larger as the nuclear and chemical incoherent scattering (cf the arrow in Fig 1-a), indicating a contribution from disordered spins, static at the experimental time scale (0.5 10$^{-11}$s). Below $T_N$, I(q) increases in the whole q range, keeping values very close to I(q)$_{HT}$ around q$\approx$0.01$\AA^{-1}$, so that a modulation appears centered at q$_m$$\approx$0.2$\AA^{-1}$ at 15K. The temperature dependence of q$_m$ cannot be detected in the 75-15K range within our accuracy. Similar observations are obtained along [112]$_1$ (mixed with  [100]$_3$), except for a small shift of 
q$_m$ towards a smaller value and a small overall decrease of intensity. I(q) along [001]$_1$ (mixed with [110]$_{2,3}$), is identical to I(q) along [110]$_1$. The same observations are obtained using an XY detector device (SANS), at 15K where the dynamical spin fluctuations can be neglected. In 
addition, in the smallest q range (q$<$0.08$\AA^{-1}$) these SANS intensities show an enhancement along [112]$_1$, which is typical of (112) interface planes between twin domains.
 
 Fig 1-b reports the magnetic contribution I$_{m}$(q) obtained by substracting I(q)$_{HT}$, for q//[110]$_1$ and q//[112]$_1$ at T=15K. The smallest q range corresponding to effects related to the dislocations, is not shown. I$_{m}$(q) consists of a broad peak, located at q$_m$$\approx 0.21$$\AA^{-1}$ and a residual and nearly flat scattering beyond 0.4$\AA^{-1}$. Along [112]$_1$, the overall intensity is slightly smaller, and a slight shift of $q_m$ is observed.

 $Elastic$ experiments
($k_i=1.55\AA^{-1}$ with Be filter), have been also performed in the ([110], [001]) plane close to $\tau$=(110)$_1$, at 8 temperatures in the $11K<T<300K$ range, and several {\bf q} directions. The [110]$_1$ q direction is reported in Fig 2-a. In this figure, the two sets of vertical lines are results of the fit of the (110)$_1$ (bold) and (002)$_2$ Bragg peaks, easily resolved due to the large orthorhombicity. At 15K, a well defined maximum in I(q) is observed at $\zeta$=0.87 (i. d. q$_m$=0.2$\AA^{-1}$), with a flat q scattering beyond. The q-symmetric intensity of this modulation with respect to (110)$_1$ is observed as a shoulder on the huge (002)$_2$ Bragg peak side of the twin domain, as shown in the inset of Fig 2-a. The apparent asymmetry of the spectrum, is due to the superposition of (002)$_2$ and (110)$_1$ Bragg peaks with no modulation from the domain (002)$_2$. We conclude that this modulation is magnetic, with a magnetization along (001). In Fig 2-b, I(q) is reported for {\bf q} along two other directions of the scattering plane at 15K, taking the {\it same} q origin at (110)$_1$. A small overall decrease of the intensity and 
a shift of q$_m$ to smaller values is observed. Along [112]$_1$, the q$_m$ value appears slightly smaller than that observed close to $\tau=0$ ( Fig 1). This is well explained by the shift of the q origin attached to the different domains, unlike the previous case, close to $\tau=0$, where the q origin is the {\it same} for all domains.
The intensity is smaller by a nearly q independent value. Actually, the intensity of the flat q tail keeps the low value determined at 300K (cf the horizontal lines in Fig 2 (a) and (b)) whereas along the other directions it varies with temperature. 
 As the temperature increases, a very slight decrease of q$_m$ can be detected whereas the intensity of the modulation decreases (Fig 2-a).  At T$_N$, a huge critical scattering intensity is observed 
close to the (110)$_1$ Bragg peak. In the T$_N$$<T<$ 250K range, small modulations appear, but along [110]$_1$ ({\bf a$_0$} axis)
 only. As shown at 250K in Fig 2-a,
the q maxima correspond to the harmonics of q$_m$. Their intensities show irreversibilities with temperature, in contrast with the broad modulation below T$_N$ which is perfectly reproducible. 
Since undetected close to $\tau$=0, they are likely related to modulated strain effects (with a Q=$\tau$+q dependent intensity) coexisting or competing with the broad magnetic modulation. At 300K, I(q) is monotonous, but its weak q dependence indicates the persistence of ferromagnetic correlations, static at the experimental
time scale. These observations confirm those obtained close to the direct beam. The flat q scattering,  with minima of intensity along [112], appears as a "sample background" corresponding to disordered spins. Although not explained, the minima of intensity along [112] could be a complementary effect of the maxima observed by SANS for $q<0.1\AA^{-1}$, and attributed to (112) interfaces between domains.   
 
We focus our analysis on the magnetic contribution I$_{m}$(q) obtained close to $\tau=0$ at 15K (Fig 1-b). A scattering  pattern showing a broad modulation with 
nearly the same q$_m$ in all {\bf q} directions, indicates a spatial organization of similar entities or " droplets", as that observed e. g.
in a chemical unmixing process\cite{Hennion3}. To apply this model in a  magnetic system, needs to define a mean magnetization density into the two phases, with a "contrast" between them, as an analog of the chemical contrast. The existence of a ferromagnetic component along {\bf c}, allows to define a magnetization function {\bf m}({\bf r}), {\bf m}$\parallel${\bf c} throughout the lattice. Therefore,
the spatial Fourier transform of the spin correlations $<S_iS_j>$ is replaced by that of $< {\bf m}({\bf r}) {\bf m}({\bf r'})>$, {\bf r} being a continuous variable in direct space. Neglecting the small anisotropy, a semi-quantitative analysis can be made using:
 
$d\sigma (q)/d\Omega =r_0^2. A.N_V.V_d^2.|\Delta m|^2.|F(qR)^2|.J(q)$                (1) 

where $\Delta m$ is the difference between the mean magnetizations {\bf m}$\parallel${\bf c} inside and outside the droplet. $F(qR)$ is the 
form factor of the droplet (a spherical shape of  radius R is assumed) of density N$_V$ and volume V$_d$ and $J(q)$ is the interference function. The A factor (A=$\Sigma _i V_i sin^2\theta _i /\Sigma _i V_i$ where $\theta _i$ is the angle between the mean magnetization {\bf m} and {\bf q} for the domain V$_i$) is required to determine $\Delta m$. It expresses that for a {\bf q} direction belonging to domain $1$, the intensity also comprises contributions from domains $2$ and $3$. They have different {\bf q} and {\bf m} directions and are weighted by their geometrical
 factor $sin^2\theta _i$. This latter factor shows that  [110] is a "pure" direction, [001] cannot be observed and 
[112] is mixed with [100], weighted by 0.5(V$_1$+V$_2$) and V$_3$ respectively. The equality of the three twinned volumes could be checked at 11K from the scattering pattern of the SANS experiment, thanks to the XY detectors device. 
We have  used the isotropic $J(q)$ function derived for liquids by Ashcroft and Lekner\cite{Ashcroft}. The spatial distribution is characterized by two parameters: the particle density $N_V$ from which we determine the  mean interparticle distance d$_m$
and the  minimal distance of approach between the particle centers $d_{min}$. The expression of
$J(q)$ is given in the footnote\cite{footnote}.  This model can fit the experimental data as shown in Fig 1-b (dashed line). It yields a diameter 2$R\approx 18\AA$ (i.e. $\approx$4-5 lattice parameter $a_0$ of the small perovskite cube),  
a droplet density $N_V=2.1\times 10^{-5}\AA^{-3}$ leading to $d_m$ $\approx 9$ $a_0$, and a minimal distance of approach $d_{min}=26\AA$ ($6\sim 7a_0$). 
This value, compared to the average distance d$_m$, accounts for the rather well defined organization and 
indicates a repulsive interaction between the droplets. 
 From the absolute value of the intensity, we determine a magnetic "contrast" 
$\Delta m\approx 0.6\pm 0.20\mu _B$.
 It corresponds to a difference of $\approx$$8$ degrees, between the canting angles characteristic of the two phases, which is surprisingly small.
  
From this semi-quantitative determination, we conclude that 
the density of the "ferromagnetic" particles is very low compared
to that of the holes (ratio 1/60 for x=0.08) leading to a picture of hole-rich droplets within a hole-poor medium\cite{Nagaev,Dagotto,Khomskii}. At x=0.08, the magnetic state of the droplet is far from a true ferromagnetic state within the canted antiferromagnetic medium. It appears canted, the spins inside the droplets being deviated from the perfect AF structure by an angle of $\approx$$18$ degrees ($\theta_{av}\approx$10 degrees). The temperature variation of the intensity below T$_N$, must be related to the evolution of 
the contrast $\Delta m$ between the two magnetic phases, which varies with the mean magnetization. The almost temperature independent evolution  
 of q$_m$ suggests that the charge carrier segregation persists at T$_N$. Above T$_N$, where the spatial distribution described above cannot be observed, the information
arises from the spin dynamics only, which indicates a ferromagnetic correlation length typical of these inhomogeneities\cite{Moussa2}.

  We compare now the characteristics of the static magnetic inhomogeneities with those of the low energy spin wave branch which appears by doping the system, considering both its q-dependent intensity and its dispersion curve.
\begin{figure}
\centerline{\epsfig{file=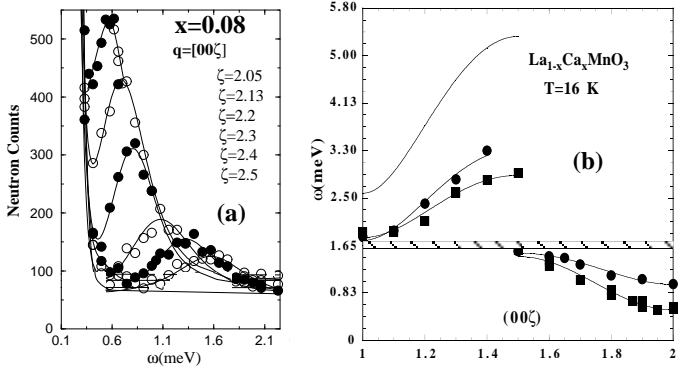,width=9cm}}
\caption{ {\bf (a)} Energy scans fitted by energy-lorentzians for q$\parallel$[$00\zeta$] at 16K. From top to bottom, $\zeta$ varies from 2.05 to 2.5.  
{\bf (b)} Dispersion curves for x=0.05 (filled circle) and x=0.08 (filled square). The continuous line at upper energy corresponds to the spin wave branch observed in pure LaMnO$_3$.}
\label{fig:35_6.8} \end{figure}
In Fig. 3 (a) we have reported some energy spectra obtained as a function of q//[001]. Similar results are observed for q//[110], indicating an isotropic and weakly dispersed spin wave branch.
As previously shown for x=0.05\cite{Hennion}, the intensity of this spin wave branch strongly decreases with q. Fitting the dynamical form factor (or energy integrated intensity) with a lorentzian function, leads to determine a ferromagnetic correlation length $\xi$ with $\xi\approx 7.5-8\AA$ for both [110] and [001]. This reveals a tight connection between this low energy spin wave branch, characteristic of a low and isotropic coupling, with the static droplets described above. 
Among other characteristic features, these spin excitations keep a propagative character and are still well-defined even at small q, i. e. over distances larger than the droplet size. These excitations may 
therefore correspond to eigenstates of the whole spin system. Within the above inhomogeneous picture derived from the static study, this means that the droplet magnetic states are coupled together through the matrix. In Fig 3-b, the dispersion curve is shown along [001] together with that 
of the other spin wave branch. The two spin wave branches observed for x=0.05 are also reported for comparison. When considering the dispersion curve of the low energy branch, a fit using $\omega$=Dq$^2$+$\omega_0$ as an approximation at small q, indicates an increase of the stiffness constant D with x, and a decrease of the gap at $\tau$=(002). Such an evolution of the constant D with the number of carriers can be expected at larger doping, within the true metallic and ferromagnetic state\cite{Kubo}. However, within the present inhomogeneous picture, we observe a very peculiar evolution of the dispersion curves with x. The corresponding two spin wave branches
 keep separated by an energy gap, constant with x (hatched area in Fig 3-b). Such a pecularity also holds at x=0.1. Therefore, the separation in the direct space into two magnetic states, induced by an electronic phase segregation is associated with a separation in the energy space into two spin dynamics, characteristic of two types of magnetic coupling. 

 In conclusion, the very unusual features found for the static and dynamical spin correlations provide new insights in the physics of the electronic phase separation. A liquid-like spatial distribution of magnetic polarons has been observed for the first time. The picture of hole-rich droplets agrees with predictions of the s-d model in a well-known limit\cite{Nagaev,Dagotto,Khomskii}. Whether such inhomogeneities exist at  x$\leq$x$_c$ where the compound becomes metallic and ferromagnetic, requires more studies. 

One author (M. H.) is very indebted to B. Hennion and D. Khomskii for fruitful discussions and G. Coddens for his critical reading of the manuscript.

\end{document}